\documentstyle [12pt,twoside]{article}

\def\frac#1#2{{\textstyle{#1\over#2}}}

\def\eg{{\it e.g.${~}$}}
\def\ie{{\it i.e.${~}$}}

\def\frac#1#2{{\textstyle{#1\over\vphantom2\smash{\raise.20ex
        \hbox{$\scriptstyle{#2}$}}}}}
\def\ha{\frac12}
\catcode`@=11
\def\underline#1{\relax\ifmmode\@@underline#1\else
        $\@@underline{\hbox{#1}}$\relax\fi}
\catcode`@=12

\def\begintitle#1#2#3#4
        {\begin{titlepage}
         \centerline{#1 \hfill UMDGR-91-#2}
         \begin{center}\vglue .7in
         {\large\bf #3}\\[.7in]
         {\bf #4}\\[.3in]
         {\it ${}^{(1)}$ Centro de Investigaciones y Estudios Avanzados,
I.P.N.}
   \\
         {\it Apdo. Postal 14-740 - Mexico 14, D.F. - Mexico}\\[.3in]
         {\it ${}^{(2)}$ Department of Physics and Astronomy}\\
         {\it University of Maryland, College Park, MD 20742}\\[.7in]
         {\bf ABSTRACT}\\
         \end{center}
         \begin{quotation}}
\def\endtitle
         {\end{quotation}
          \end{titlepage}
          \newpage}

\begin{document}

\begintitle{January 1992}{134}{REMARKS ON PURE SPIN CONNECTION
FORMULATIONS OF GRAVITY}{Riccardo
Capovilla${}^{(1)}$ and Ted Jacobson${}^{(2)}$}
In the derivation of a pure spin connection action functional
for gravity two methods have been proposed. The first starts
from a first order lagrangian formulation, the second from
a hamiltonian formulation. In this note we show that
they lead to identical results.

\vspace{1cm}

PACS: 04.20.Cv, 04.20.Fy, 02.40.+m

\endtitle

As shown by Ashtekar, there is a hamiltonian formulation of
general relativity which uses the (spatial) self-dual spin connection
as the configuration space variable.\cite{Ashtekar1}
The 3-metric is essentially the conjugate momentum, and therefore loses
some of its priviledged status in the theory. It is possible to go one step
further, and give a covariant lagrangian formulation of
general relativity which employs only the (space-time) self-dual
spin connection and a scalar density as the gravitational
variables.\cite{CDJ1,CDJ3} In this ``pure spin connection formulation",
the spacetime metric
is merely a derived quantity, and general relativity appears as a generally
covariant gauge theory of an SL(2,C) connection.

For a spacetime with a metric of lorentzian signature, the use of the
self-dual spin connection as a field variable implies the use of
{\it complex} variables. Properly speaking therefore the
pure spin connection formulation describes complex general
relativity, and reality conditions need to be imposed a posteriori.
For the sake of simplicity, in the following we will consider only the
complex case. The appropriate reality conditions are discussed
in detail in Ref. \cite{CDJ3}.

The pure spin connection formalism was originally derived starting from a
first order formulation of general relativity, and solving the variational
equations in favor of the self-dual spin connection, an undetermined
scalar density, and the matter fields.\cite{CDJ1,CDJ3}
It was later shown that one can also arrive there directly from Ashtekar's
hamiltonian formalism, by performing an inverse Legendre
transform.\cite{BP1,Peldan1,Peldan2}
In the vacuum case the pure spin connection action turns out to be polynomial
in the basic field variables. For massless scalar  and
spinor field couplings, it remains polynomial. However in the case
of a non-vanishing cosmological constant, or for massive scalar or
gauge field couplings, it becomes non-polynomial and rather complicated.

A caveat which needs to be mentioned is that the
equivalence between the usual formulation of general relativity
and its pure spin connection version holds only modulo  a certain
non-degeneracy condition. The form of this condition
depends on the type of matter coupling considered, and on the
presence of a cosmological constant. In the vacuum
case, for a vanishing cosmological constant,
the condition is that the self-dual part of the Weyl
tensor, thought as three by three matrix, must be non-degenerate.
This implies a restriction on the class of spacetimes for which the
equivalence holds. In particular, it fails for spacetimes of type
$\{ 3,1\}$, $\{ 4\}$ in the algebraic classification of the Weyl
tensor.\cite{PR2}

A discrepancy appeared in the non-polynomial cases between the results obtained
by the two approaches mentioned above. It turns out that in these cases there
was an error in the derivation of a pure spin connection
action functional of \cite{CDJ3}.\cite{CJErratum} In this note
we shall show that in fact the two approaches lead to identical results.

In order to
demonstrate the equivalence we have found it useful to slightly rewrite
the first order formulation of general relativity that
serves as the starting point in the derivation of a pure spin-connection
action. In particular, we shall write the constraint on the trace
of the Lagrange multiplier matrix field $X$
(which in the vacuum case turns out to correspond to the self-dual part of
the Weyl tensor)
in a form alternative to the obvious choice Tr$X = \Lambda$ used in
\cite{CDJ3}.
Dadhich {\it et al.} have previously pointed out that
different ways of imposing this constraint may lead to
different pure spin-connection actions. \cite{Dadhich}
Such differing actions are
related by field redefinition of the scalar density that appears in the final
form. In the polynomial cases, the result turns out to
be identical to what was obtained before.

We now go on to the calculations. First we rederive the pure spin-connection
action in the vacuum case. This rederivation turns out to be simpler
than the original derivation given in \cite{CDJ3}, because an awkward step
requiring the extraction of the square root of a $3\times3$ matrix is
eliminated. In fact it is in this step that the error was made in the
cosmological constant case.\cite{CJErratum}

\vspace{4mm}

Complex vacuum general relativity can be described by the
first order action functional introduced originally by Plebanski
\cite{Plebanski2} (see also \cite{MF}, \cite{SD2F}). This action
is a functional of a trio of 2-forms $\Sigma^i$,
an SO(3,C) connection 1-form $\omega^i$, with curvature $R_i := d\omega_i +
\frac{1}{2} \epsilon_{ijk} \omega^j \wedge \omega^k$, and a
symmetric and traceless Lagrange multiplier field
$\Psi_{ij}$\footnote{Lower
case latin letters from the middle of the alphabet denote SO(3,C)
indices, which are raised and lowered with the Kronecker delta
$\delta^{ij}$.}.

The action can be written in the form
\begin{equation}
S [\Sigma^i ,\omega_i , \Psi_{ij} , \mu ] = \int \, [\, \Sigma^i \wedge R_i
- \frac{1}{2} \Psi_{ij} \Sigma^i \wedge \Sigma^j + \mu \Psi^i{}_i \, ]
\label{eq:1}
\end{equation}
Let us briefly discuss the variational equations for this action.
(For a more detailed discussion see \eg \cite{SD2F}.)
The Lagrange multiplier $\mu $ enforces explicitly the condition that
$\Psi_{ij}$ is traceless. In turn, the $\Psi_{ij}$ equations of motion
imply that the trio of
2-forms $\Sigma^i$ are subject to the constraint
$ \Sigma^i \wedge \Sigma^j \propto \delta^{ij} $.
The content of this constraint is that there exists a tetrad of 1-forms
$\theta^a$ such that the trio of 2-forms corresponds to the self-dual part
of $\theta^a\wedge\theta^b$. The trio of
2-forms $\Sigma^i $ play the role of basic ``metric variables".
The $\omega_i $ field equation
implies then that $\omega_i $ itself is the self-dual part of the spin
connection compatible with the tetrad $\theta^a $. Finally, the $\Sigma^i $
equation of motion, $ R_i = \Psi_{ij} \Sigma^j $,
says that the curvature is pure Weyl, \ie that
the spacetime metric  $ g_{\mu\nu}= \theta_\mu{}^a \theta_\nu{}^b \eta_{ab}$
is Ricci flat.
The metric density can also be expressed directly in terms
of the 2-forms $\Sigma^i $ as
\begin{displaymath}
\sqrt{g} g_{\mu \nu} = \frac{1}{3} \; \Sigma_{\mu\alpha}^i
\Sigma_{\beta\gamma}^j \Sigma_{\delta\nu}^k\;
\epsilon^{\alpha\beta\gamma\delta}\epsilon_{ijk}.
\end{displaymath}
In this sense one can say that $\Sigma^i$ is a``cube root" of the
metric.

Starting from the action (\ref{eq:1}), it is possible to eliminate
both $\Sigma^i $
and $\Psi_{ij} $, solving for them in terms of the curvature
2-form $ R_i $ and an undetermined scalar density by use of the
variational equations \cite{CDJ1,CDJ3}.
This results in a pure spin connection formulation of (complex)
vacuum general relativity given by
\begin{equation}
 S [\omega, \eta ] = \int \eta [\mbox{Tr} (M^2) - \frac{1}{2} (\mbox{Tr} M)^2 ]
\label{eq:2}
\end{equation}
where we have defined the matrix (of weight $+ 1$)
\begin{equation}
M_{ij} := R_i \wedge R_j
\label{eq:dm}
\end{equation}
and $\eta $ is a scalar density of weight $- 1$.
In terms of $\eta$ and $\omega_i$, the spacetime metric is given by
$$g_{\mu\nu} = \frac{1}{3} \; \eta\; R_{\mu\alpha}^i
R_{\beta\gamma}^j R_{\delta\nu}^k\;
\epsilon^{\alpha\beta\gamma\delta}\epsilon_{ijk}.$$
{}From the point of view of the pure connection
formulation, the way to identify the metric is to
express the $\omega_i$ equation of motion as $D\Sigma^i=0$
for some 2-forms $\Sigma^i$. These $\Sigma^i$ are then the cube root
of the metric density. They are determined up to a constant multiplicative
factor.

The equivalence between the actions (\ref{eq:1}) and (\ref{eq:2})
holds if and only if
$\Psi_{ij}$ is invertible. The reason is that this assumption is necessary
in order to solve the $\Sigma $ variational equation,
$R_i = \Psi_{ij} \Sigma^j$,
for $\Sigma^i $ itself, \ie $ \Sigma^i = (\Psi^{-1})^{ij} R_j$.
This implies the restriction mentioned previously on the class of spacetimes
for which the pure spin-connection formulation is equivalent to general
relativity.

We shall now derive the pure spin-connection action (\ref{eq:2}), and its
generalization
with a cosmological constant, from the first order action (\ref{eq:1}).
In preparation,
the constraint on Tr$\Psi$ will first be rewritten. This is done with the help
of the characteristic equation satisfied by any $3 \times 3$ matrix $A$, \ie
\begin{equation}
A^3 - (\mbox{Tr}A)A^2+ \frac{1}{2} ((\mbox{Tr} A)^2-\mbox{Tr} A^2)A -
\mbox{det} A = 0
\label{eq:ch}
\end{equation}
When $\Psi$ is non degenerate, which we need to assume anyway, we can
take $A=\Psi^{-1}$, multiply through by $\Psi$, and take the trace to obtain
\begin{equation}
\mbox{Tr} \Psi = \frac{1}{2}\, \mbox{det} \Psi\;
[  (\mbox{Tr} \Psi^{-1})^2 - \mbox{Tr} \Psi^{-2} ]
\label{eq:cond}
\end{equation}
With this alternative expression for $\mbox{Tr} \Psi$, the action (\ref{eq:1})
takes the form
\begin{equation}
S [\Sigma ,\omega , \Psi , \rho ] = \int \mbox{Tr} [\Sigma \wedge R
- \frac{1}{2} \Psi \Sigma \wedge \Sigma ] +
\rho [(\mbox{Tr} \Psi^{-1})^2 - \mbox{Tr} \Psi^{-2} ]
\label{eq:3}
\end{equation}
where the Lagrange multiplier $\mu$ has been traded in for $\rho$,
which is related to $\mu$ by $\rho = \mu(\mbox{det} \Psi)/2$.
(The indices have been suppressed in a convenient ``matrix" notation.)

The $\Sigma$ variational equation  is $ R = \Psi \Sigma $, which is solved by
$\Sigma = \Psi^{-1} R$, provided that $\Psi$ is invertible.
Since one has solved for the
same variable $\Sigma $ that was varied in the action, the solution can
be substituted back into the action, yielding
\begin{equation}
 S[\omega, \Psi, \rho ]  =    \int \frac{1}{2}\mbox{Tr} [\j^{-1} M ] +
\rho [(\mbox{Tr} \Psi^{-1})^2 - \mbox{Tr} \Psi^{-2} ]
\label{eq:4}
\end{equation}
where $M$ is defined as in (3).
At this point one wants to solve for $\Psi$ from the $\Psi$ equation of motion,
and here the alternative form of the trace constraint turns out to be
more convenient. The variation of (\ref{eq:4}) with respect to $\Psi^{-1}$
gives
\begin{displaymath}
 M = 4\rho [ \Psi^{-1} - (\mbox{Tr} \Psi^{-1}) I ]
\end{displaymath}
which is easily solved for $\Psi^{-1}$ with respect to $ M$ and $\rho$, \ie
\begin{displaymath}
\Psi^{-1} = (4 \rho )^{-1} [ M - \frac{1}{2} (\mbox{Tr} M) I ]
\end{displaymath}
When this solution is substituted in the action (\ref{eq:4}), one recovers
the action (\ref{eq:2}), with $\eta=(16\rho)^{-1}$.

One may wonder what happens if in the constraint term in the action
(\ref{eq:3}), the
relative coefficient $- 1$ is replaced by an arbitrary parameter.
A derivation which follows the footsteps of the one just given for vacuum
general relativity shows that the elimination of $\Sigma$ and $\Psi$
recovers the family of generally covariant gauge theories
discussed in \cite{Capovilla} which generalize  the pure
spin connection action for vacuum general relativity (see also \cite{BPIJMP}).

\vspace{4mm}

We now turn to the case of a non-vanishing cosmological constant.
The addition of a cosmological constant term to (\ref{eq:1})
yields the action
\begin{equation}
  S[\Sigma,\omega, X, \mu ] =
  \int \{ \mbox{Tr} [\Sigma \wedge R - \frac{1}{2} X
    \Sigma \wedge \Sigma ] + \mu(\mbox{Tr}X-\L) \}
\label{eq:cc}
\end{equation}
where we have defined $X := \Psi + (1/3) \Lambda I$,
and $\Lambda$ is the cosmological constant. The Lagrange multiplier $\mu$
enforces the constraint that $X$ has trace equal to $\Lambda$, or, in
other words, that $\Psi$ is traceless.

Provided that the assumption of invertibility of X holds, (\ref{eq:cond})
can be used to replace the constraint $\mbox{Tr} X -\Lambda=0$ in the action
(\ref{eq:cc}) by the equivalent constraint
\begin{equation}
(\mbox{Tr} X^{-1})^2 - (\mbox{Tr} X^{-2}) - 2\, (\mbox{det}X)^{-1} \L = 0
\label{eq:c2}
\end{equation}
This yields the action
\begin{equation}
  \int \{ \mbox{Tr} [\Sigma \wedge R - \frac{1}{2} X
    \Sigma \wedge \Sigma ] + \rho \; [\, (\mbox{Tr X}^{-1})^2 -
(\mbox{Tr} X^{-2}) - 2\, (\mbox{det}X)^{-1} \Lambda \, ] \, \}
\label{eq:cc1}
\end{equation}
where now $\mu$ has been traded for $\rho = \mu(\mbox{det} X)/2$.

Solving the $\Sigma $ equation of motion for $\Sigma$ itself gives
\begin{equation}
 \int \{ \frac{1}{2} \mbox{Tr}(X^{-1}  M) + \rho \;
[\, (\mbox{Tr X}^{-1})^2 - (\mbox{Tr} X^{-2})
- 2\, (\mbox{det}X)^{-1} \Lambda \, ] \, \}
\label{eq:5}
\end{equation}
where $M$ is defined as in (\ref{eq:dm}). Now varying this action
(\ref{eq:5}) with respect to $X^{-1}$ gives
\begin{equation}
M = 4 \rho [ X^{-1} - (\mbox{Tr}X^{-1}) {\bf I}
+ \Lambda (\mbox{det} X)^{-1} X]
\label{eq:emme}
\end{equation}
which is a matrix equation quadratic in $X^{-1}$. (That the last term is
quadratic in $X^{-1}$ follows from the characteristic equation (\ref{eq:ch})
with
$A=X^{-1}$.) If this equation could be solved for $X$ in terms of $M$ and
$\rho$, the solution $X(M,\rho)$ could then be substituted back into the action
(\ref{eq:5}), yielding an action involving only the spin connection and $\rho$.

Now in fact it does {\it not} seem to be possible to obtain a solution for
$X(M,\rho)$ in closed form. Moreover, such a solution may be non-unique or may
not exist at all for certain values of $(M,\rho)$. (For instance, the
simpler equation $M=B^2$ has many solutions for $B$ if $M=I$ and {\it none}
if all elements of $M$ vanish except for $M_{12}=M_{33}=1$.)

Peld\'an also faces the problem of solving the matrix equation (12)
in his alternative approach. \cite{Peldan2}
His method of dealing with this situation is the following. One supposes there
is a unique solution $Y= Y(M, \rho):= X^{-1}(M,\rho) $ to (\ref{eq:emme}).
One then makes the move of substituting the expression (\ref{eq:emme}) for $M$
into  the action (\ref{eq:cc}), obtaining
\begin{equation}
\int \rho
[\mbox{Tr} Y^2 - (\mbox{Tr}Y)^2  + 4\Lambda (\mbox{det} Y)]
\label{eq:y}
\end{equation}
This action involves only the particular combination of invariants
\begin{equation}
z=z(\mbox{Tr}M,\mbox{Tr}M^2,\mbox{det}M,\rho):=
\mbox{Tr} Y^2 - (\mbox{Tr}Y)^2  + 4\Lambda (\mbox{det} Y)
\end{equation}
Although one cannot solve (\ref{eq:emme}) explicitly for $Y=X^{-1}$,
it turns out that one {\it can}
solve for the invariant $z$. Is this sufficient? It can happen that
the matrix equation (\ref{eq:emme}) has no solution but that there is
a solution for its invariants. Ignoring these pathological cases,
to solve for $z$, consider
the independent invariants of the equation (\ref{eq:emme}),
\begin{eqnarray}
\mbox{Tr}\; M &=&  2 \rho \; \{\, \Lambda \, [( \mbox{Tr}Y)^2
- (\mbox{Tr}Y^2)] - 4 (\mbox{Tr}Y)\} \nonumber \\
\mbox{Tr}M^2 &=& 16 \rho^2 \{ (\mbox{Tr} Y^2)
+ (\mbox{Tr} Y)^2
- \Lambda (\mbox{Tr} Y)[ (\mbox{Tr} Y)^2
- (\mbox{Tr} Y^2) ] +  6 \Lambda (\mbox{det} Y)
\nonumber\\
  &-& 2 \Lambda^2 (\mbox{Tr} Y) (\mbox{det} Y)  + \frac{1}{4}
\Lambda^2 [(\mbox{Tr} Y)^2
- (\mbox{Tr} Y^2)]^2 \} \nonumber \\
\mbox{det} M &=& 64 \rho^3 \{ (\mbox{det} Y) - \ha  (\mbox{Tr} Y)
[ (\mbox{Tr} Y)^2 -
 (\mbox{Tr} Y^2)] + \frac{1}{4} \Lambda [ (\mbox{Tr} Y)^2 -
(\mbox{Tr} Y^2)]^2 \nonumber \\
&+& \Lambda  (\mbox{Tr} Y) (\mbox{det} Y) - \Lambda^2 (\mbox{det} Y)
[ (\mbox{Tr} Y)^2 -
 (\mbox{Tr} Y^2)]
+ \Lambda^3 (\mbox{det} Y)^2 \} \nonumber
\end{eqnarray}
(The last expression is obtained by computing Tr$M^3$ and then
using the characteristic equation (\ref{eq:ch}) for $M$.)
Although we have not been able to solve these equations for
$\mbox{Tr}Y$, $\mbox{Tr} Y^2$ and $\mbox{det} Y$ independently,
one can in fact solve for the combination $z$. It turns out that it
obeys the quadratic equation
\begin{equation}
{\Lambda^2 \over 8} z^2 - (1 + {\Lambda \over 8 \rho} \mbox{Tr}M) z
+ {1 \over 16 \rho^2} [ \mbox{Tr}M^2 - \frac{1}{2} (\mbox{Tr} M)^2 ]
- {\Lambda \over 32 \rho^3}  \mbox{det} M = 0
\end{equation}
The integrand
in (\ref{eq:y}) can then be written in terms of $M$ and $\rho$ as follows
\begin{eqnarray}
S &=& {1 \over 2 \Lambda} \int d^4 \, x \chi^{-1} \;
[\; ( 1 + \chi \, \mbox{Tr}M) \nonumber \\
& & \pm \{( 1 + \chi \, \mbox{Tr}M)^2 -  2 \chi^2 \,
[ \mbox{Tr} M^2 - \ha (\mbox{Tr}M)^2 ]
+  8 \chi^3 \, (\mbox{det} M) \}^{1/2} \; ]
\label{eq:peldan}
\end{eqnarray}
where $\chi:=(8\rho )^{-1} \Lambda$. This is exactly the action found
by Peld\'an in \cite{Peldan2}.

There are {\it two} distinct actions, depending on which square root
is taken in (\ref{eq:peldan}). This is a consequence
of the fact that (\ref{eq:emme}) is non-linear in the unknown $X^{-1}$.
Both roots are needed, since omitting a root would
arbitrarily exclude some field configurations that satisfy the equations of
motion\footnote{Note that we disagree here with Peld\'an \cite{Peldan2}, who
argues that one should keep only the root that
agrees with the vacuum action (\ref{eq:2}) in the limit of a vanishing
$\Lambda$.}.

What if the same procedure used for the action (\ref{eq:cc1}) is applied
to the equivalent action (\ref{eq:cc}), that is without rewriting the
constraint
Tr$X = \Lambda$ in the form (\ref{eq:c2})? It turns out that although
the X equation of motion, $M = 2 \rho X^2$, is simpler than (\ref{eq:emme}),
it is harder to solve for the
invariant Tr$X$ that appears in the Lagrangian.  The independent traces
of the equation $M = 2 \rho X^2$ can be used to obtain a quartic equation
for Tr$X$, but we have been unable to obtain an explicit
expression for its solutions.

\vspace{1cm}

\noindent ACKOWLEDGEMENTS

\vspace{.3cm}

We thank Abhijit Kshirsagar for a useful discussion. This work was
supported by NSF grants PHY 8910226 and PHY 9112240.
R.C. gratefully acknowledges support by a CONACyT post-doctoral fellowship,
and by the Centro de Investigaciones y Estudios Avanzados, I.P.N. (Mexico).

\end{document}